\documentclass[twocolumn,preprintnumbers,
superscriptaddress,amsmath,floatfix]{revtex4}

\usepackage{amssymb}  
\usepackage{graphicx,subfigure}
\usepackage{exscale}
\usepackage{textcomp}
\usepackage{enumerate}
\usepackage{amsmath}
\usepackage{amssymb}
\usepackage{color,graphicx}
\usepackage[dvipsnames]{xcolor}
\usepackage{bbm}
\usepackage[pdfstartview=XYZ,
bookmarks=true,
colorlinks=true,
linkcolor=blue,
urlcolor=blue,
citecolor=blue,
pdftex,
bookmarks=true,
linktocpage=true, 
hyperindex=true
]{hyperref}
\usepackage{orcidlink}

\usepackage{color}

\usepackage[normalem]{ulem}  

\newcommand{\nsat}{n_0}

\newcommand\eps{\varepsilon}

\newcommand\s{\sigma}
\newcommand\si{\sigma}

\newcommand\om{\omega}

\newcommand{\be}{\begin{equation}}
\newcommand{\ee}{\end{equation}}
\newcommand{\bea}{\begin{eqnarray}}
\newcommand{\eea}{\end{eqnarray}}
\newcommand{\ba}[1]{\begin{array}{#1}}
\newcommand{\ea}{\end{array}}

\newcommand{\bm}[1]{\mbox{\boldmath${#1}$}}

\newcommand{\Msolar}{\ensuremath{{\rm M}_\odot}}

\newcommand{\eqn}[1]{(\ref{#1})}
\newcommand{\sliver}{\kern 0.07em} 

\newcommand{\chiPT}{\raisebox{0.2ex}{\ensuremath{\chi}}\text{EFT}}
\newcommand{\chiEFT}{\raisebox{0.2ex}{\ensuremath{\chi}}\text{EFT}}
\newcommand{\meff}{M_*}
\newcommand{\half}{{\textstyle\frac{1}{2}}}
\newcommand{\MeV}{\text{MeV}}
\newcommand{\fm}{\text{fm}}
\newcommand{\km}{\text{km}}

\newcommand{\sev}{\langle\sigma\rangle}
\newcommand{\wev}{\langle\omega_0\rangle}
\newcommand{\rev}{\langle\rho_{03}\rangle}

\newenvironment{tightlist}[2]{ 
\begin{list}{#2}{
  \usecounter{#1}
  \setlength{\topsep}{0ex} 
  \setlength{\itemsep}{-\parsep} 
  \settowidth{\labelwidth}{#2} 
  \setlength{\labelsep}{0.2em}        
  \setlength{\leftmargin}{\labelwidth}
  \addtolength{\leftmargin}{\labelsep}
 }}{\end{list}}

\begin{document}
\preprint{LA-UR-22-23972}

\title{Relativistic mean-field theories for neutron-star physics based on chiral effective field theory}

 \author{M. G. Alford \,\orcidlink{0000-0001-9675-7005}
}
 \email{alford@physics.wustl.edu}
 \affiliation{Department of Physics, Washington University in St.~Louis, St.~Louis, MO 63130, USA}
 
 \author{L. Brodie\,\orcidlink{0000-0001-7708-2073
}}
 \email{b.liam@wustl.edu}
 \affiliation{Department of Physics, Washington University in St.~Louis, St.~Louis, MO 63130, USA}
 
 \author{A. Haber\,\orcidlink{0000-0002-5511-9565}}
 \email{ahaber@physics.wustl.edu}
 \affiliation{Department of Physics, Washington University in St.~Louis, St.~Louis, MO 63130, USA}

\author{I. Tews\,\orcidlink{0000-0003-2656-6355
}}
\email{itews@lanl.gov}
\affiliation{Theoretical Division, Los Alamos National Laboratory, Los Alamos, NM 87545, USA}
\date{18 July 2022}

\begin{abstract}
We describe and implement a procedure for determining the couplings of a Relativistic Mean-Field Theory (RMFT) that is optimized for application to neutron star phenomenology.
In the standard RMFT approach, the couplings are constrained by comparing the theory's predictions for symmetric matter at saturation density with measured nuclear properties.
The theory is then applied to neutron stars which consist of neutron-rich matter at densities ranging up to several times saturation density, which allows for additional astrophysical constraints.
In our approach, rather than using the RMFT to extrapolate from symmetric to neutron-rich matter and from finite-sized nuclei to uniform matter, we fit the RMFT to properties of uniform pure neutron matter obtained from chiral effective field theory.
Chiral effective field theory incorporates the experimental data for nuclei in the framework of a controlled expansion for nuclear forces valid at nuclear densities and enables us to account for theoretical uncertainties when fitting the RMFT.
We construct four simple RMFTs that span the uncertainties provided by chiral effective field theory for neutron matter, and are consistent with current astrophysical constraints on the equation of state. 
Our RMFTs can be used to model the properties of neutron-rich matter across the vast range of densities and temperatures encountered in neutron stars and their mergers.
\end{abstract}

\maketitle

\section{Introduction}
\label{sec:intro}

Relativistic mean-field theories (RMFTs) can be used to  
model nuclear matter across the range of densities and temperatures found in neutron stars and neutron star mergers \cite{Walecka1974,Boguta:1977xi,Serot:1984ey,Glendenning1996}. 
Importantly, RMFTs can in principle be used to calculate any well-defined physical property of neutron star matter.  
In addition to the equation of state (EoS) and its physical temperature dependence~\cite{Serot:1984ey},~RMFTs provide a consistent framework for, e.g., out-of-equilibrium behavior~\cite{Alford:2021ogv}, response to magnetic fields~\cite{Broderick:2000pe,Haber:2014ula}, 
and the spectrum of low-energy excitations, which is needed for calculations of phenomenologically-relevant properties such as bulk viscosity~\cite{Alford:2019qtm,Alford:2019kdw}, neutrino opacity and emissivity~\cite{Roberts:2016mwj}, etc.
These properties of neutron star matter are key inputs for robust simulations of explosive scenarios involving neutron stars, such as supernova explosions and neutron star mergers.

Matter in neutron stars is highly neutron rich, but 
such systems are not self-bound, hence there are no direct experimental data 
that could be used to determine appropriate values for the couplings of an RMFT in this regime. 
RMFTs are, therefore, generally calibrated by fitting the couplings to data extrapolated from experimental results for atomic nuclei using the Bethe-Weizs{\"a}cker liquid-drop model, i.e., to the properties of \emph{symmetric} nuclear matter (equal numbers of protons and neutrons) around nuclear saturation density ($\nsat$) or ground state properties of \emph{nearly symmetric} finite nuclei, e.g., charge radii and binding energies of closed-shell nuclei. The RMFT is then  
extrapolated to neutron star conditions, where matter is far from symmetric, with a proton fraction below $10\%$. 
In addition, to describe neutron stars, RMFTs need to be extrapolated from nuclear densities to neutron star densities of up to several times $\nsat$. In principle, the symmetry energy and its slope at $\nsat$ provide some information on the extrapolation to neutron-rich matter, but, as highlighted by the recent Lead Radius Experiment (PREX-II) results, their values remains highly uncertain~\cite{Adhikari:2021prex,Reed:2021prex_slope_sym}.
This further motivates finding alternative ways of determining coupling constants of RMFTs aimed at describing neutron-rich matter probed in neutron stars. See Ref.~\cite{Hornick:2018kfi} for an RMFT parametrization that checks for agreement with chiral effective field theory (\chiEFT) for neutron matter.

\begin{figure}
    \centering
    \includegraphics[width=1\hsize]{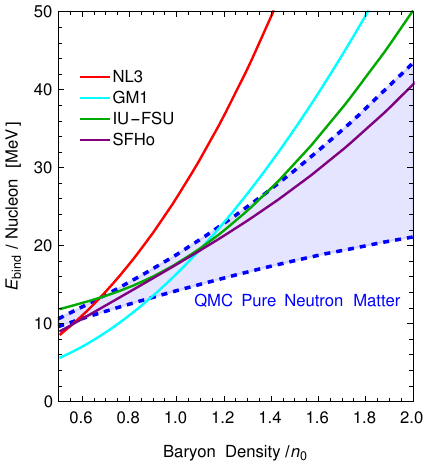}
    \caption{
       Binding energy per nucleon as a function of baryon number density in pure neutron matter.
    A comparison between the GM1~\cite{GM1}, NL3~\cite{Lalazissis:1996rd}, IU-FSU~\cite{IUF}, and SFHo~\cite{Steiner:2012rk} EoSs and the \chiPT\ uncertainty band of Ref.~\cite{Tews:2018kmu}.}
    \label{fig:iuf_chipt}
\end{figure}

In this paper, we construct a set of RMFTs that are particularly well suited to describing neutron-rich matter.
We constrain the RMFT couplings using the best information we have about
neutron-rich matter, which comes from \chiEFT\ calculations for pure neutron matter. In Fig.~\ref{fig:iuf_chipt} we show the \chiPT\ uncertainty band for the binding energy of neutron matter, along with the predicted values from some commonly used RMFTs that are calibrated directly to symmetric matter. We see that they all show some level of disagreement with \chiPT. 

The advantage of \chiEFT\ is that it provides a controlled expansion that describes the interactions between nucleonic degrees of freedom in the density range around $\nsat$ consistent with all symmetries of the fundamental theory of strong interactions, Quantum Chromodynamics (QCD)~\cite{Epelbaum:2008ga,Machleidt:2011zz}.
Because of that, \chiEFT\  calculations enable us to make reliable connections between the near-symmetric matter probed in experiments with atomic nuclei and neutron-rich matter found in neutron stars~\cite{Lynn:2015jua,Lonardoni:2019ypg,Drischler:2020yad}.
Here, we determine the couplings in our new RMFTs by fitting to the \chiEFT\  predictions for zero-temperature neutron matter reported in Ref.~\cite{Tews:2018kmu}. 
We use these results over the density range $0.5\nsat$ to $2\nsat$, above the crust-core transition and where \chiEFT\  was found to be reliable~\cite{Tews:2018kmu,Drischler:2020yad}. We observe that commonly used RMFTs (see Fig.~\ref{fig:iuf_chipt}) are inconsistent with \chiEFT\  at densities less than $1.5\nsat$ or even well below saturation.
At densities above $2\nsat$, \chiEFT\  breaks down because nucleon momenta approach the breakdown scale of the theory, implying that additional degrees of freedom become important.
The \chiEFT\  results employed in this work are obtained by solving the nuclear many-body problem with quantum Monte Carlo (QMC) many-body methods~\cite{Carlson:2014vla}, which are among the most precise many-body methods in nuclear physics.
These calculations provide ``synthetic'' data for neutron-rich matter that can be used to adjust RMFTs.
Moreover, we simultaneously require that the RMFTs reproduce the basic properties of symmetric nuclear matter, namely the saturation density, binding energy, and incompressibility at that density.

The resulting RMFTs are used to extrapolate the EoS to higher densities, beyond the range of validity of \chiEFT, 
and to model the behavior of neutron-rich matter found in neutron stars.  
The RMFT couplings can then be further constrained by comparing predictions for neutron star properties with astrophysical measurements, such as radio~\cite{Antoniadis:2013pzd,Cromartie:2019kug} or x-ray~\cite{Miller:2021qha,Riley:2021pdl} observations of masses and radii of heavy neutron stars, 
and extractions of the tidal deformability from gravitational-wave (GW) observations, e.g., GW170817~\cite{TheLIGOScientific:2017qsa,Abbott:2018gw170817}.

The result of our work is a family of RMFTs that are well suited for neutron star phenomenology; they can be used in calculations of r-mode spindown and neutron star cooling, as well as in neutron star merger simulations. In these simulations, the matter is generally neutron-rich but may explore a range of proton fractions as it is driven out of beta equilibrium in the first milliseconds of the merger.
We provide our new EoSs in a publicly available tabular form on the CompOSE website~\url{https://compose.obspm.fr/eos/277} and the code needed to generate the tables and figures in a GitLab repository: \url{https://gitlab.com/ahaber/qmc-rmfx}.

This paper is organized in the following way.
In Sec.~\ref{sec:models} we
describe the \chiEFT\  calculation of the low-density neutron-matter EoS, and the procedure that we follow to fit an RMFT to that data.
In Sec.~\ref{sec:results} we present our main results,
showing key physical properties of a sample of resulting RMFTs.
Sec.~\ref{sec:conclusions} gives our conclusions.
In all our calculations we use natural units, \mbox{$\hbar=c=k_B=1$}.

\section{Microscopic Models}
\label{sec:models}

\subsection{Chiral Effective Field Theory for Pure Neutron Matter}
\label{sec:chiPT}

The properties of nucleonic matter are determined by the strong nuclear interactions between neutrons and protons.
The fundamental theory for these interactions is QCD, which describes nuclear systems in terms of quark and gluon degrees of freedom.
It is currently unfeasible to describe nucleonic matter in terms of these degrees of freedom because QCD is nonperturbative in this regime.
However, the relevant low-energy degrees of freedom in nuclear physics are nucleons, neutrons and protons.
\chiEFT\ is a low-energy effective field theory of QCD that describes strong interactions in terms of nucleonic degrees of freedom~\cite{Epelbaum:2008ga,Machleidt:2011zz}.
It derives nuclear Hamiltonians from the most general Lagrangian consistent with all the symmetries of QCD which is expanded in powers of momenta over the breakdown scale $\Lambda_b$, leading to a systematic expansion that is truncated at a chosen order.

\chiEFT\ explicitly includes pion-exchange interactions, leading to a breakdown scale of the order of the $\rho$-meson mass. 
Recent numerical studies found that the breakdown scale lies in the range $500-600$~MeV~\cite{Drischler:2020yad}.
Physics at higher energies is encoded in general contact interactions that depend on low-energy couplings (LECs) fit to nucleon-nucleon (NN) scattering data and data on atomic nuclei. 
At leading order (LO), nuclear Hamiltonians from \chiEFT\ include momentum-independent contact interactions, describing NN $S$-wave scattering, and the well known one-pion-exchange interaction. 
By going to higher orders in the effective field theory, additional pion-exchange and contact interactions are accounted for, leading to more accurate and precise calculations at the cost of more complicated Hamiltonians.
Because of the systematic expansion provided by \chiEFT, it is possible to estimate theoretical uncertainties from order-by-order calculations~\cite{Epelbaum:2014efa,Drischler:2020hwi,Drischler:2020yad}. 
This is one of the main benefits of \chiEFT\ over other approaches to dense nuclear matter. 
Furthermore, \chiEFT\ naturally accounts for three-nucleon (3N) and higher many-body forces, and these forces are consistent with the NN sector for the interactions, i.e., the same physical processes are described by the same operators and LECs.
\chiEFT\ provides a natural hierarchy for NN and 3N forces, with 3N interactions starting to contribute at next-to-next-to-leading order (N$^2$LO).

In this work, we employ the \chiEFT\ calculations of pure neutron matter reported in Ref.~\cite{Tews:2018kmu}.
These calculations used local \chiEFT\ interactions at N$^2$LO that were constructed in Refs.~\cite{Gezerlis:2013ipa,Gezerlis:2014zia,Tews:2015ufa,Lynn:2015jua}.
The NN interactions were fit to NN scattering data and the 3N interactions to the binding energy of $^4$He and neutron-$^4$He scattering phase shifts.
These interactions were used in QMC computations of neutron matter using the auxiliary field diffusion Monte Carlo (AFDMC) method~\cite{Schmidt:1999lik}.
AFDMC calculations with local chiral interactions have been shown to reproduce properties of dense matter~\cite{Lonardoni:2019ypg,Tews:2018kmu} and atomic nuclei~\cite{Lynn:2015jua,Lonardoni:2017hgs} with great success.
The neutron-matter results of of Ref.~\cite{Tews:2018kmu} include theoretical uncertainties from several sources: 
(a)~the truncation of the \chiEFT\ series, estimated following Ref.~\cite{Epelbaum:2014efa},
(b)~uncertainties due to the choice of employing local regulators~\cite{Lynn:2015jua,Huth:2017wzw}, and (c)~stochastical uncertainties from the AFDMC many-body method. 
The latter source of uncertainty is negligible and the theoretical error is dominated by the truncation and regulator uncertainties.
The pure neutron-matter results employed here are based in the TPE+$V_{\rm E\mathbbm{1}}$ and TPE-only results of Ref.~\cite{Tews:2018kmu} with their uncertainties.

\subsection{Relativistic Mean-Field Model}
\label{sec:RMFT}

We now show that simple, generic RMFTs can reproduce the nuclear matter properties
that are most relevant to neutron star applications. 
Our RMFT Lagrangians are based on Appendix A of Ref.~\cite{Alford:2021ogv}, and include
protons and neutrons coupled to $\si$, $\om$, and $\rho$ mesons. By setting selected couplings to zero, these Lagrangians can be reduced to the form of well known RMFTs like GM1~\cite{GM1}, IU-FSU~\cite{IUF}, or NL3~\cite{Lalazissis:1996rd}. In the future, as new data becomes available or the \chiPT\ uncertainty bands tighten, there might be good reason to explore more complicated RMFTs with additional mesonic fields and interactions.

Our RMFT Lagrangian can be written as a sum of nucleonic, mesonic, photon, and leptonic parts, 

\begin{align}
    \mathcal{L}=\mathcal{L}_{\text{N}}
    +\mathcal{L}_{\text{M}}
    +\mathcal{L}_{\pi}
    +\mathcal{L}_l 
    +\mathcal{L}_\gamma 
    + B\, ,
    \label{eq:Lagrangian}
\end{align}
where we separate the pionic part as explained later in this section.
The ``bag constant,'' $B$, arises because this RMFT may not accurately model very low density physics ($n_B \ll \nsat$), such as formation of clusters or nuclei, so its vacuum state may have a nonzero pressure (in the convention that the physical vacuum has zero pressure).
When constructing neutron star density profiles, we will determine $B$ by matching to a low-pressure EoS (see Sec.~\ref{sec:crust}) at $n_B \approx 0.40\nsat$.

The nucleonic Lagrangian $\mathcal{L}_{\text{N}}$ describes the neutrons and protons, combined into an isodoublet, $\psi$, and their Yukawa-type coupling to the mesonic fields (apart from the pion), 
\begin{align}
    \mathcal{L}_{\text{N}}&=\Bar{\psi}\big[i\gamma^{\mu}\partial_{\mu}-m_N
    +g_{\sigma}\sigma  -g_{\omega}\gamma^{\mu}\omega_{\mu}\nonumber\\
    &-\frac{g_{\rho}}{2}\boldsymbol{\tau}\cdot\boldsymbol{\rho}_{\mu}\gamma^{\mu}+\frac{e}{2}(1+\tau_3)\gamma^{\mu}A_{\mu}\big]\psi\, ,
\end{align}
where bold symbols denote isovectors and $\boldsymbol{\tau}$ is the isovector of isospin generators. The last term describes the photon-nucleon coupling, with $e$ being the electrical charge, $A^\mu$ is the $U(1)_e$ gauge field, and where we use the third isospin matrix $\tau_3$ to construct the isospin projector for the proton.

The pion Lagrangian $\mathcal{L}_\pi$ contains pion-nucleon and pion-pion interactions. 
Although pions play a major role in long-range nucleon-nucleon interactions, they only play a minor role in bulk properties of nuclear matter because a pion expectation value would break parity; see, for example, p.~494 in Ref.~\cite{Walecka1974}.
Hence, they do not influence the astrophysical properties that we use to constrain our RMFTs, and play no role in our analysis.

The Lagrangian for the nonpionic mesons is given by
\begin{align}
    \mathcal{L}_{\text{M}} &= 
    \frac{1}{2}\partial_{\mu}\sigma\partial^{\mu}\sigma-\frac{1}{2}m^2_{\sigma}\sigma^2
    -\frac{bM}{3}(g_{\sigma}\sigma)^3-\frac{c}{4}(g_{\sigma}\sigma)^4 \nonumber \\
    & -\frac{1}{4}\omega_{\mu\nu}\omega^{\mu\nu}+\frac{1}{2}m^2_{\omega}\omega_{\mu}\omega^{\mu} +\frac{\zeta}{24}g^4_{\omega}(\omega_{\mu}\omega^{\mu})^2 \nonumber\\
 & -\frac{1}{4}\boldsymbol{B}_{\mu\nu}\cdot\boldsymbol{B}^{\mu\nu}+\frac{1}{2}m^2_{\rho}\boldsymbol{\rho}_{\mu}\cdot\boldsymbol{\rho}^{\mu}+b_{1}g^2_{\rho}\omega_{\nu}\omega^{\nu}\boldsymbol{\rho}_{\mu}\cdot\boldsymbol{\rho}^{\mu} \,,
\label{eq:mesons}
\end{align}
where
\begin{align}
    \omega_{\mu\nu}\equiv\partial_{\mu}\omega_{\nu}-\partial_{\nu}\omega_{\mu}\,,\\
    \boldsymbol{B}_{\mu\nu}\equiv\partial_{\mu}\boldsymbol{\rho}_{\nu}-\partial_{\nu}\boldsymbol{\rho}_{\mu}\,.
\end{align}
The leptonic Lagrangian is
\begin{equation}
 \mathcal{L}_l=\bar{\psi}_e\left(i\gamma^\mu\partial_\mu-m_e\right)\psi_e\,,
\end{equation} 
describing free electrons with mass $m_e=0.511\,\MeV$. In principle, muons could be added to the theory as another noninteracting free lepton field, but they play a subleading role relative to the electrons in establishing electrical neutrality of bulk nuclear matter and affect the total pressure at the one percent level. Although the proton fraction can change from roughly $10\%$ up to $15\%$ compared to the case without muons, the smaller Fermi momentum of the electrons counterbalances this effect. The resulting effect on the direct Urca threshold density is therefore only at the few percent level. 

The photon contribution is given by
\begin{equation}
    \mathcal{L}_\gamma=-\frac{1}{4}F^{\mu\nu}F_{\mu\nu}\, ,
\end{equation}
with the standard electromagnetic field-strength tensor $F^{\mu\nu}$.
The parameters of the theory are determined as follows:
the three meson masses are fixed to the values used in the IU-FSU RMFT, $m_{\sigma} = 
  491.5\,\MeV$, \mbox{$m_{\omega} = 782.5\,\MeV$},
$m_{\rho} = 763.0\,\MeV$ \cite{IUF}.
The parameter $M$ is an arbitrary constant with units of mass which is introduced so that the $\si^3$ coupling $b$ in Eq.~\eqref{eq:mesons} will be dimensionless, and is set to the vacuum nucleon mass \mbox{$M=m_N=939$ MeV}.
The $\om$ meson quartic self-coupling $\zeta$ is poorly constrained in the density region of our fit, and can lead to unphysical behavior of the theory \cite{Mueller:1996pm} if the coupling is negative. Furthermore, its primary purpose is to soften the EoS; see Ref.~\cite{Todd-Rutel:2005yzo}. 
Given that our models predict reasonable maximum masses and in order to avoid unphysical models due to an unconstrained parameter (varying $\zeta$ after a successful fit does not alter the quality of the fit to a significant degree)  we set $\zeta$ to zero in this work.
The remaining six couplings, $(g_\si, g_\omega, g_\rho, b, c, b_1)$ are obtained by fitting to data as described in Sec.~\ref{sec:fitting}.

In the mean-field approximation, we treat the meson fields as classical, with translationally and rotationally invariant expectation values which are obtained by maximizing the pressure (see Appendix \ref{appx:eom}).  
Because of rotational invariance, only the zeroth component of the Lorentz four-vector fields $\om$ and $\rho$ can acquire an expectation value. Only the third (charge-neutral) isospin component of the $\bm{\rho}_0$ isovector acquires an expectation value (see p.~184 in Ref.~\cite{Glendenning1996}).
The resulting dispersion relations for the nucleons $i=n,p$ are 
\begin{equation}
      E_i=\sqrt{k_i^2+M_*^2}+g_{\omega}\langle\omega_0\rangle+g_{\rho}I_{i3}\langle\rho_{03}\rangle \, , \label{eq:disprel} 
\end{equation}
  with
  \begin{equation}
        \meff = m_N-g_{\sigma}\langle\sigma\rangle \, ,
    \label{eq:meff}
\end{equation}
and where $I_{i3}=-\half$ or $+\half$ for $n$, $p$ respectively.
We see that the coupling to the $\si$ field produces an effective mass $\meff$, the coupling to the $\om$ produces an energy shift common to both nucleon species, and the coupling to $\rho$ creates opposite energy shifts for the neutron and proton.

The RMFT can be used to calculate physical observables at any temperature, but for comparison with current neutron star observations and the fit to the \chiEFT\  data we only need the low temperature ($T\ll 1\,\MeV$) behavior. 
Given that the nucleon 
Fermi energies are much larger than the temperature in isolated neutron stars, we will use the $T=0$ approximation for comparisons with data. 
In Appendix~\ref{appx:eom}, we give finite-temperature expressions, and show thermal pressures. 

The pressure can be decomposed as follows:

\begin{align}\label{eq:totpress}
    P=\langle \mathcal{L}_{\text{M}} \rangle+P_N+P_e+B \, .
\end{align}
The first term represents the mesonic contribution and is obtained directly from the Lagrangian by replacing all fields with their expectation values. In the mean-field approximation, this part is temperature independent. In the zero-temperature limit, the nucleonic contribution $P_N$ can be obtained analytically (see Sec.~3 in Ref.~\cite{Schmitt:2010pn}) and is given by the pressure of free nucleons obeying the dispersion relations \eqn{eq:disprel},
\begin{align}
P_N =\sum_{i=n,p}\frac{1}{8\pi^2} & \biggl[\left(\frac{2}{3}k_{Fi}^3-\meff^2 k_{Fi}\right)E_{Fi}^* \nonumber \\
&+ \meff^4\ln\frac{k_{Fi}+E^*_{Fi}}{\meff}\biggr]\, ,   
\label{eq:Pnuc}
\end{align}
where we defined the effective Fermi energy
\begin{equation}
    E_{Fi}^*=
    \mu_i-g_{\omega}\langle\omega_0\rangle-g_{\rho}I_{i3}\langle\rho_{03}\rangle=\sqrt{k_{Fi}^2+\meff^2} \, .
    \label{eq:estar}
\end{equation}
The Fermi momenta are connected to the baryon densities in the usual way,
\begin{equation}
 n_{i}=\frac{k_{Fi}^3}{3\pi^2} \, .  
  \label{eq:nb}
\end{equation}
The third term in Eq.~(\ref{eq:totpress}) is the free electron contribution to the pressure, where we impose local charge neutrality of the system by demanding that the magnitude of the electron Fermi momentum is equal to the proton Fermi momentum, $k_{Fe}=k_{Fp}$.
We obtain the expectation values of the meson fields by maximizing the pressure (see Appendix \ref{appx:eom} for details),
\begin{align}\label{eq:P-max}
\frac{\partial P}{\partial\langle\sigma\rangle}=0 \, , \qquad \frac{\partial P}{\partial\langle\omega_0\rangle}=0 \, , \qquad \frac{\partial P}{\partial\langle\rho_{03}\rangle}=0 \, .
\end{align}

\subsection{Fitting a Relativistic Mean-Field Theory to Chiral Effective Field Theory}
\label{sec:fitting}

As described in Sec.~\ref{sec:intro},  our goal is  to derive RMFTs that are tuned to the properties of neutron-rich matter.
We therefore constrain the six free parameters $(g_\si, g_\omega, g_\rho, b, c, b_1)$ of our RMFTs by fitting the models to the following data:
\newcounter{counta}
\begin{tightlist}{counta}{$\bullet$}
  \item the \chiEFT\  EoS of zero-temperature homogeneous neutron matter in the density range where \chiPT\ is expected to be reliable ($0.5\nsat$ to $2\nsat$);
  \item the saturation density $\nsat$, binding energy at $\nsat$, and incompressibility of symmetric nuclear matter;
  \item the astrophysical constraint that the EoS should at least be able to support the mass of the heaviest observed neutron star.
\end{tightlist}

\smallskip
Our procedure is as follows:
\begin{enumerate}
  \item To sample the \chiPT\ uncertainty band, we create a set of representative \chiEFT\  neutron matter EoSs using the Gandolfi-Carlson-Reddy (GCR) parametrization developed in Ref.~\cite{Gandolfi:2011xu} which expresses the binding energy per nucleon $\mathcal{E}$ of pure neutron matter in the form
  \begin{equation}
    \mathcal{E}_{\chiPT}(n_{B}) = a (n_{B}/\nsat)^\alpha + b (n_{B}/\nsat)^\beta.
    \label{eq:GCR}
  \end{equation}
  To create the set, we sample a range of values of the GCR parameters $(a,b,\alpha,\beta)$, keeping only the samples that remain within the \chiEFT\  uncertainty band (Fig.~\ref{fig:iuf_chipt}) for the energy per particle and within the \chiEFT\ uncertainty range for the pressure at $2n_0$.
  \item For each representative $\chiPT$ EoS, we perform a nonlinear least-squares fit to find the set of RMFT couplings that lead to the best agreement with the $\chiPT$ binding energy $\mathcal{E}(n_{B})$ for neutron matter at $0.5\nsat$ to $2\nsat$, and also with the standard values for properties of symmetric nuclear matter at nuclear saturation density (see details below). 
  \item The resultant RMFTs can be used to obtain an EoS for cold beta-equilibrated nuclear matter \mbox{($\mu_n=\mu_p+\mu_e$)} from about $0.5\nsat$ up to several times $\nsat$. At densities up to $0.5\nsat$ we use a crust EoS as described in Sec.~\ref{sec:crust}. We can then obtain predictions for the mass-radius relation and tidal deformability of neutron stars. We keep the RMFTs whose $M(R)$ relation has a maximum mass that reaches $2\si$ compatibility with the heaviest known neutron stars (see details below) and discard the others.
  \end{enumerate}
By repeating the above procedure with progressively finer sampling of the GCR parameter space, we are
able to find RMFTs that agree with the \chiEFT\  predictions for neutron matter, the basic properties of symmetric matter at saturation density, and are consistent with astrophysical data. 

\medskip
We now give additional technical details about the fitting (step 2) and the astrophysical constraints (step 3).
Our final results, including a set of RMFTs that meet all our criteria, are presented in Sec.~\ref{sec:results}.

\medskip
\noindent\underline{Step 2: Fitting to nuclear matter properties}.\\[0.4ex]
We fit to the following nuclear matter data:

(a) The binding energy of isospin-symmetric matter in the range of densities $(0.8\nsat, 1.4\nsat)$. We use 12 sample points in this range, at which we evaluate the binding energy using the standard empirical power series around saturation density,
\begin{equation}
    \mathcal{E}(n_B,\alpha)=(B_\text{sat}+\frac{\kappa}{2!}\delta^2+\cdots)+\alpha^{2}(J+L\delta+\cdots)+\cdots,
    \label{eq:snm_power_series}
\end{equation}
where $\delta\equiv(n_{B}-\nsat)/(3\nsat)$ and $\alpha\equiv(n_n-n_p)/(n_n+n_p)$; $\alpha$ represents the asymmetry of the matter, where $n_n$ and $n_p$ are the neutron and proton number densities, respectively.
In symmetric nuclear matter $\alpha=0$.
$B_\text{sat}$ is the binding energy at saturation density, and $\kappa$ is the incompressibility of nuclear matter.
The parameter values that we fit to are
\begin{subequations}
\begin{align}
  B_\text{sat}^\text{expt} &= -16\,\MeV,\\
  \nsat^\text{expt} &= 0.16\,\fm^{-3},\\
  \kappa^\text{expt} &= 240\pm20\,\MeV,
  \end{align} \label{eq:sym-matter-expt}
\end{subequations} where the binding energy is extracted in the traditional approach of extrapolating the semiempirical mass formula~\cite{Myers:1995wx}.
We note that other methods have been proposed that lead to slightly less binding, \mbox{$B_\text{sat}=-(13-14)$}~MeV~\cite{Atkinson:2020yyo}.
We use the standard value for saturation density from 
Ref.~\cite{Bethe:1971xm}, which is still consistent with more recent estimates, $0.15\pm0.01$~\cite{Horowitz:2020evx}, and the incompressibility from Ref.~\cite{Shlomo:2006incompressibility}.
Our choices are also consistent with the values adopted in Ref.~\cite{Huth:2020ozf}.
%

(b) A representative \chiEFT\  neutron-matter EoS, which is defined via Eq.~\eqn{eq:GCR} by a set of values of $(a,b,\alpha,\beta)$. We sample the binding energy $\mathcal{E}(n_B)$  at 16 density values in the range $(0.5\nsat, 2\nsat)$. 

For each representative \chiEFT\  EoS, we use the multiple nonlinear model fitting function available for {\em  Mathematica} \cite{wolframrep} to find the set of RMFT couplings $(g_\si, g_\omega, g_\rho, b, c, b_1)$ that best reproduce the data described in (a) and (b) previously. The RMFT's predictions for the binding energy are calculated using the thermodynamic relation at vanishing temperature \mbox{$\varepsilon=-P+\sum_i\mu_i n_{i}$} with $i=n,p,e$. To perform a fit and evaluate its accuracy one must associate error bars with the data points. We explored several different recipes for this: $10$\% error bars on all data points; $1\,\MeV$ error bars on all data points; and $10$\% error bars on all data points for pure neutron matter and a varying error between $5-10$\% for symmetric nuclear matter, where the error grows away from saturation density. Comparing the different results, we find that the fits are not sensitive to the choice of error bar. For example, the saturation properties of isospin-symmetric matter change by less than $1$\%. The RMFTs that achieve a sufficiently good fit (coefficient of determination $R^2>0.99 $) are then screened for astrophysical validity (step 3).

\medskip
\noindent \underline{Step 3: Constraint from astrophysical data.} \\[0.4ex]
For each set of RMFT couplings that successfully reproduced the nuclear matter data as described in step 2 above, we calculated the EoS, i.e., the energy density as a function of pressure, $\eps(P)$, for homogeneous matter in beta equilibrium at zero temperature. The assumption of homogeneous matter is only valid down to about $0.4\nsat$, and below that density we switch to the model described in Sec.~\ref{sec:crust}. 
We use this combined EoS to solve the Tolman-Oppenheimer-Volkoff (TOV) equations to calculate the neutron star mass-radius curve. 
We discard RMFTs for which the maximum mass is inconsistent at the $2\si$ level with the best constraint on the heaviest neutron star, $M = (2.072\pm0.066)\Msolar$ \cite{Riley:2021pdl}.

The code used to generate these representative $\chiPT$ EoSs, RMFT fits, and resulting mass-radius curves is available at \url{https://gitlab.com/ahaber/qmc-rmfx}.

\subsection{Attaching the Crust}
\label{sec:crust}

In principle, one can use a homogeneous RMFT calculation to describe nuclear matter down to the vacuum. However,
at densities below about $0.4\nsat$, where the matter begins to clump into ``pasta'' structures, we would have to drop our assumption that the field expectation values are translationally invariant, and switch to a more complex calculation using Wigner-Seitz cells in which the matter is described self-consistently in the Thomas-Fermi approximation.
This procedure is too computationally demanding to be performed for every set of RMFT couplings that we consider.
We therefore use the GPPVA(TM1e) crustal EoS from CompOSE \cite{compose:gppva_tm1e}. This EoS combines the Baym-Pethick-Sutherland (BPS) EoS \cite{Baym:1971pw} for the outer crust below a density of $n_B=0.002\, \mathrm{fm}^{-3}$ with a Thomas-Fermi calculation \cite{Grill:2014aea} using the TM1e RMFT \cite{Shen:2020sec} for the inner crust. 
We match this combined crustal EoS to the EoS for our RMFT by demanding thermodynamic consistency (continuity in pressure and baryon chemical potential and monotonicity of the baryon density with respect to the baryon chemical potential) at crust baryon density $n_\mathrm{tr}\approx 0.4\nsat$. For each of our RMFTs, we impose a first-order phase transition at the highest density in the GPPVA(TM1e) crust that ensures thermodynamic consistency. 
To accomplish the matching, we tune the bag constant in our RMFT in Eq.~(\ref{eq:Lagrangian}), which penalizes the pressure of the homogeneous nuclear matter phase relative to the crust phase.

\section{Results}
\label{sec:results}

\begin{table*}
\begin{tabular}{lccccccccccc}
\hline
Name & $g_\si$ & $g_\om$ & $g_\rho$ & $b$ & $c$ & $b_1$ & $B$ & $n_\text{sat}$ & $\mathcal{E}(n_\text{sat})$ & $\kappa(n_\text{sat})$  & $M_\text{max}$ \\[-0.3ex]
    Unit &     &     &     &     &     &   & [MeV$^4$] & [$\fm^{-3}$] & [MeV] & [MeV] & [$\Msolar$] \\ 
\hline
QMC-RMF1 & 7.54 & 8.43 & 10.88 & 0.0073 & 0.0035 & 7.89 & -612215 & 0.160 & -16.1 & 260 & 1.95  \\
QMC-RMF2 & 7.82 & 8.99 & 11.24 & 0.0063 & -0.0009 & 8.02 & -463438 & 0.161 & -16.3 & 264 & 2.04  \\
QMC-RMF3 & 8.32 & 9.76 & 11.02 & 0.0063 & -0.006 & 5.87 & -707480 & 0.157 & -16.1 & 230 & 2.15  \\
QMC-RMF4 & 8.21 & 9.94 & 12.18 & 0.0041 & -0.0021 & 10.43 &  -206742 & 0.162 & -16.1 & 279 & 2.21  \\
\hline
\end{tabular}
\caption{
Couplings and fitted properties for four selected RMFTs that meet the criteria described in Sec.~\ref{sec:fitting}.  
We find that these theories predict properties of isospin-symmetric matter that are close to the inferred values in Eq.~\eqn{eq:sym-matter-expt}, and their
predicted maximum neutron star masses are consistent with observations. 
}
\label{tab:couplings}
\end{table*}

\begin{table*}
\begin{tabular}{lcccccccccc}
\hline
Name & $J$ & $L$ & $M_*(n_\text{sat})/m_N$ & $R_{1.4\Msolar}$ & $\Lambda_{1.4\Msolar}$ & $n_{B} (M_{\text{max}})$ \\[-0.3ex]
    Unit & [MeV] & [MeV] &  & [km] &  & [$\nsat$]\\ 
\hline
QMC-RMF1 & 32.9 & 44.5 & 0.782 & 11.86& 313 & 7.1\\
QMC-RMF2 & 32.7 & 40.6 & 0.759 & 12.03 & 357 & 7.0 \\
QMC-RMF3 & 33.6 & 49.2 & 0.732 & 12.26 & 387 & 6.8\\
QMC-RMF4 & 30.4 & 31.3 & 0.716 & 12.35 & 470 & 6.2\\
Inference/observation 
 & 29-32 
 & 40-65
&
 & $12.45\pm0.65$ & $190^{+390}_{-120}$ &  \\
\hline
\end{tabular}
\caption{
Selected properties of the four RMFTs in Table~\ref{tab:couplings}.  
Inferred ranges for $J$ and $L$ come from Fig.~2 (the white ``intersection'' region) in Ref.~\cite{Drischler:2020hwi},
for the neutron star radius from Ref.~\cite{Miller:2021qha}, and for the tidal deformability from Ref.~\cite{Abbott:2018gw170817}. 
}
\label{tab:predictions}
\end{table*}

\begin{figure}
    \centering
    \includegraphics[width=1\hsize]{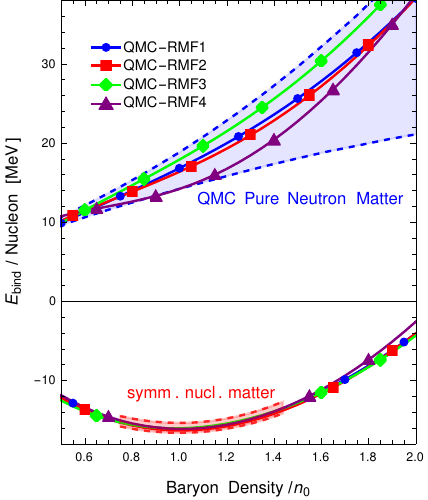}
    \caption{Binding energy per nucleon as a function of baryon number density in pure neutron matter and isospin-symmetric nuclear matter. The blue shaded region represents the uncertainty band of the QMC \chiPT\ calculation. The red shaded region shows the symmetric matter values \eqn{eq:sym-matter-expt} that we fit to. The solid lines are results for the four RMFTs obtained in Sec.~\ref{sec:fitting}. Note that the symbols on the lines are distinguishing markings, not data points.
    }
    \label{fig:pnm_snm_eb}
\end{figure}

\begin{figure}
    \centering{
    \includegraphics[width=1\hsize]{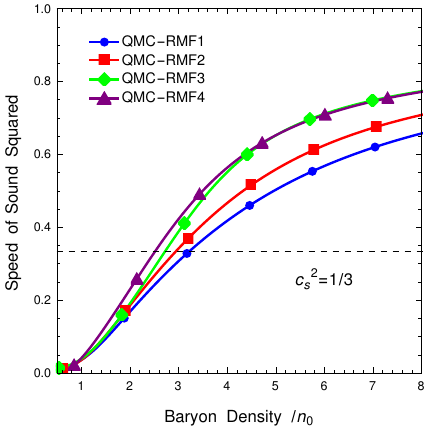}
    }
    \caption{Speed of sound squared in beta-equilibrated nuclear ($npe$) matter as a function of baryon number density in units of the speed of light, for the four  RMFTs obtained in Sec.~\ref{sec:fitting}. Note that the symbols on the lines are distinguishing markings, not data points. The dashed line at $c_{s}^{2}=1/3$ shows the speed of sound in the QCD asymptotic high-density limit. 
       }
    \label{fig:speed_of_sound}
\end{figure}

We selected four representative RMFTs that emerged from the fitting process described in Sec.~\ref{sec:fitting} that cover a range of stiffness, from the softest (QMC-RMF1) to the stiffest (QMC-RMF4).
Table~\ref{tab:couplings} lists the fitted values of the couplings for those RMFTs. 

Each of these EoS predicts energies per particle for pure neutron matter 
that lie within the QMC \chiPT\ uncertainty band.
This is illustrated in Fig.~\ref{fig:pnm_snm_eb}, where we show the energy per particle as a function of density for neutron matter (upper panel) and isospin-symmetric matter (lower panel) for the four RMFTs listed in Table~\ref{tab:couplings}. 

We note that at $n_B>\nsat$ all RMFTs lie in the upper part of the \chiPT\ uncertainty band, because the astrophysical requirement of a $\approx 2\Msolar$ star requires a sufficiently high pressure at those densities, which is proportional to the slope of the energy per particle.
We also note that all four RMFTs  agree well with the expected features of isospin-symmetric matter at densities close to $\nsat$ (red shaded region in the bottom panel of Fig.~\ref{fig:pnm_snm_eb}); see also Table~\ref{tab:couplings}. 
Moreover, each RMFT can support neutron stars with masses in agreement with the heaviest observed star. 
The strongest lower bound on $M_\text{max}^\text{expt}$,  $(2.072\pm0.066)\Msolar$, was extracted by the NICER Collaboration~\cite{Riley:2021pdl}, and the predicted mass-radius curve must reach at least $1.94\,\Msolar$ to be compatible with observations at the $2\si$ level.

In Fig.~\ref{fig:speed_of_sound} we show the speed of sound $c_s^2=dP/d\eps$ as a function of density in beta-equilibrated homogeneous $npe$ matter described by our four RMFTs.
As expected for a relativistic theory, the speed of sound remains causal, i.e., it is always less than the speed of light. 
Our four RMFTs are named by their stiffness in the density range 2-4$\nsat$, which is the typical central density of  $1.4\,\Msolar$ stars, and therefore influences their radius.
The QMC-RMF1 theory produces the softest (lowest sound speed) EoS and QMC-RMF4 producing the stiffest EoS, although above $4\,\nsat$ its speed of sound is very similar to QMC-RMF3.
For all RMFTs, the speed of sound rises above the conformal value $c_s^2=1/3$ at about $3\,\nsat$.
This is consistent with more generic analyses (e.g., Refs.~\cite{Bedaque:2014sqa,Tews:2018kmu,Tan:2020ics,Drischler:2021bup}) which indicate that this is a necessary feature of any EoS that is consistent with nuclear-physics information at low densities and also meets the astrophysical constraint $M_\text{max}\gtrsim 2\,\Msolar$.

For all four EoSs, the speed of sound increases monotonically, rising quickly at $n_B\approx 2\nsat$ and then leveling off, asymptotically approaching the speed of light at high densities. One might expect that in the limit of infinite density the theory would become scale free ($\mu_B$ much greater than all other physical energy scales) in which case 
$c_s^2=1/3$.  However, in our theory the $\omega$ expectation value dominates the pressure at high density, so the theory does not become scale invariant. For a discussion of this point, see Appendix~\ref{appx:csasym}. In a more general theory with a non-zero $\om^4$ coupling
one would find $c_s^2\to1/3$ in the infinite density limit.

\begin{figure}
    \centering{
    \includegraphics[width=1\hsize]{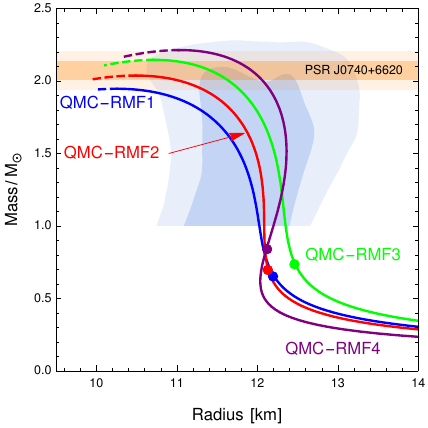}
    }
    \caption{Mass-radius curves for the RMFTs obtained in Sec.~\ref{sec:fitting}. The orange-shaded bars show the $1\si$ (dark shading) and $2\si$ (light shading) mass measurement of pulsar J0740+6620 from Ref.~\cite{Riley:2021pdl}. 
    The blue shaded area shows the 68\% (dark shading) and 95\% (light shading) multimessenger constraints obtained in Ref.~\cite{Pang:2021jta} using the NICER+XMM-Newton result of Miller et al.~\cite{Miller:2021qha}.
    The dots mark where the central density rises above $2\nsat$ and our models are no longer constrained by \chiPT\. 
    The dashed parts beyond the maxima of the curves denote the unstable branch of solutions. }
    \label{fig:MR_max_mass}
\end{figure}

Fig.~\ref{fig:MR_max_mass} shows the mass-radius relations predicted by the proposed RMFTs and we indicate where the central density reaches $2\nsat$.
Around this density, \chiPT\ results become unreliable and other theoretical approaches, such as an RMFT, are needed to model the microphysical properties of the matter. 
Hadronic theories like RMFTs become implausible at densities around $6\nsat$ because nucleons start to overlap.
From Table~\ref{tab:predictions} we find that the maximum baryon number density attained in stars is comparable to this value, so a neutron star spans the range of densities over which an RMFT is applicable.
The figure shows that, as expected by their construction, all four RMFTs are consistent with the maximum mass constraint. 
We also show (blue shaded region) the multimessenger constraint on mass and radius obtained in Ref.~\cite{Pang:2021jta} combining observations of two radio pulsars, two NICER measurements (using the NICER and XMM-Newton result of Ref.~\cite{Miller:2021qha} for J0740+6620), the gravitational-wave detections GW170817 and GW190425 and modeling of the associated kilonova AT2017gfo and the gamma-ray burst GRB170817A. 
The mass-radius relations predicted by our RMFTs are consistent with that multimessenger constraint.

The mass-radius curves are also consistent with the stiffness hierarchy that we observed in the speed of sound (Fig.~\ref{fig:speed_of_sound}). 
QMC-RMF1 is the softest EoS in the 2-4$\nsat$ range, and correspondingly yields the most compact stars and the smallest maximum mass.
QMC-RMF2 is stiffer, with larger radii and a higher maximum mass.
The QMC-RMF4 theory produces the stiffest EoS in the 2-4$\nsat$ range, so that the radius increases with mass in the range from about 0.6 to $1.5\,\Msolar$, which corresponds to central densities from $1.7\nsat$ to $2.9\nsat$.
As shown in Table~\ref{tab:predictions}, these theories give radii for a typical $1.4\,\Msolar$ star and tidal deformabilities that are consistent with current observations, e.g., Refs.~\cite{Dietrich:2020lps,Al-Mamun:2020vzu,Miller:2021qha,Raaijmakers:2021uju,Legred:2021hdx,Pang:2022rzc}. The effective nucleon masses $M_*(n_{\text{sat}})$ in our RMFTs (see Table~\ref{tab:predictions}) are consistent with the values found by other microscopic calculations in the literature \cite{Bodmer:1989hdx,Margueron:2017eqc,Li:2018lpy,Chen:2014mza}. We find that the effective mass 
has an inverse correlation with 
the maximum mass and the radius of a $1.4\,\Msolar$ star, with a sensitivity
similar to that found in   Refs.~\cite{Hornick:2018kfi,Ghosh:2021bvw}. Additionally, we have verified that our models are consistent with the perturbative QCD constraints from  Ref.~\cite{Gorda:2022jvk}.

Table~\ref{tab:predictions} also shows what our RMFTs predict for the standard parameters of isospin-dependence of the nuclear matter EoS at saturation density, namely the parameters $J$ and $L$ [see Eq.~\eqn{eq:snm_power_series}]. The symmetry energy $J$ is the difference between the binding energy per nucleon of neutron matter and symmetric nuclear matter at $\nsat$. The predicted values from our RMFTs for $J$, $30$ to $34$ MeV, are consistent with constraints from Refs.~\cite{Lattimer:2012xj,Lattimer:2014sga,Drischler:2020hwi}.
The slope of the symmetry energy $L$ characterizes how the symmetry energy varies with density. The values predicted by our RMFTs are in the $30$ to $50\,\MeV$ range which is consistent (with some tension for QMC-RMF4) with the constraints displayed as the white ``intersection'' area in Ref.~\cite{Drischler:2020hwi}, Fig.~2, based on the analyses of
Refs.~\cite{Lattimer:2012xj,Lattimer:2014sga}.

They are also consistent at the $2\si$ level with the mean values extracted from the recent PREX–II experiment using covariant energy density functionals~\cite{Reed:2021prex_slope_sym}, $L = 106(37)\,\MeV$. 

Another property that is relevant to neutron star phenomenology, specifically cooling, is the direct Urca threshold, which is the density at which the proton fraction becomes high enough so that the Urca process $n\to p+e^{-}+\bar\nu_e$ can  proceed without in-medium corrections.
For all our proposed RMFTs, we find the direct Urca threshold to be at high densities, far above the value $n_B\approx 6\nsat \approx 1\,\fm^{-3}$, where, as noted above, the RMFT is no longer a plausible description as nucleons start to overlap. 
This means that in order to explain a fast cooling scenario for the heaviest neutron stars one would need to explore more complicated RMFT Lagrangians with additional fields or couplings, or supplement the RMFTs with a phase transition to a quark matter phase that supports direct Urca processes \cite{Grigorian:2004jq,Heinke:2008vj,Negreiros:2010tf,Noda:2011ag,Sedrakian:2015qxa,deCarvalho:2015lpa}.

\section{Conclusions}
\label{sec:conclusions}

In this paper, we described a procedure for obtaining RMFTs that are particularly suited to the description of neutron star matter.  
An RMFT provides a full microscopic description of the physics explored in neutron stars.
In addition to the temperature and density dependent EoS, RMFTs can in principle provide predictions for any physical observable, including out-of-equilibrium behavior, transport properties, etc.
Because neutron star matter is neutron-rich, we constrained the RMFT couplings by fitting to  ``data'' for pure neutron matter that is obtained indirectly, via experimentally constrained \chiPT\ calculations. 
In addition, we required that 
our RMFTs agree well with the known properties of isospin-symmetric nuclear matter near saturation density, and are consistent with astrophysical constraints on the mass-radius relation of neutron stars.

Using this procedure we obtained four representative RMFTs with varying stiffness that can be used to provide a full description of neutron-rich matter at densities where \chiPT\ is inapplicable.
These RMFTs are designed to model nuclear matter across the range of densities and temperatures found in neutron stars and neutron star mergers. 
Our RMFTs predict that $1.4\,\Msolar$ neutron stars will have radii ranging from $11.8$ to $12.3\,\km$, which is consistent with observations~\cite{Dietrich:2020lps,Riley:2021pdl}.

The procedure described here can readily be repeated for (a) different (e.g.,~higher-order) \chiPT\ calculations, (b) different or additional constraints on masses, radii, or other neutron star properties, and (c) for more complicated RMFT Lagrangians.
For example, we were able to achieve good agreement with current data using a simple RMFT with six variable couplings, but these models can be expanded by adding more meson fields and more couplings.

\bigskip

{\centering\bf Acknowledgements \par}
We thank Sophia Han, Jorge Piekarewicz, J\"urgen Schaffner-Bielich, Andreas Schmitt, and Ziyuan Zhang for their input. 
This research was partly supported by the U.S. Department of Energy, Office of Science, Office of Nuclear Physics, under Award No.~\#DE-FG02-05ER41375,
and performed in part at Aspen Center for Physics, which is supported by National Science Foundation Grant No. PHY-1607611.
The work of I.T. was supported by the U.S. Department of Energy, Office of Science, Office of Nuclear Physics, under Contract No.~DE-AC52-06NA25396, by the Laboratory Directed Research and Development program of Los Alamos National Laboratory under Project No. 20220658ER, and by the U.S. Department of Energy, Office of Science, Office of Advanced Scientific Computing Research, Scientific Discovery through the Advanced Computing (SciDAC) NUCLEI program.
\appendix

\section{Pressure and Field Equations}
\label{appx:eom}
\begin{figure}
    \centering{
    \includegraphics[width=1\hsize]{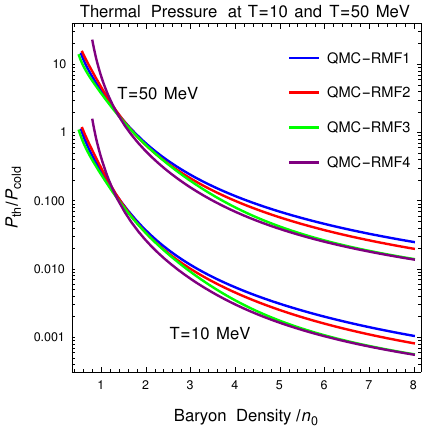}
    \caption{Relative contribution of the thermal pressure to the pressure at $T=0$ for all four models at $T=10$ MeV (lower set of curves) and $T=50$ MeV (upper set of curves).    }
    \label{fig:PfinteT}}
\end{figure}

In this appendix we present the finite temperature field equations that determine the meson field expectation values of our RMFT. 
For neutron star phenomenology the relevant densities (and baryon chemical potentials) are so large that we can safely neglect antibaryons (for the full treatment including them, see, e.g.,~\cite{Schmitt:2010pn}).
The finite temperature nucleon pressure is given by
 \begin{equation}
     P_N=2T\sum_{i=n,p}\int\frac{d^3k}{(2\pi)^3}\ln\left[1+e^{-\left( E_i-\mu_i\right)/T}\right] \, ,
 \end{equation}
 where $E_i$ are the dispersion relations given in Eq.~(\ref{eq:disprel}). The meson contribution to the pressure remains unchanged at finite temperatures, and the electron pressure is given by the pressure of a free relativistic Fermi gas, where we can neglect the electron mass compared to its chemical potential,
 \begin{equation}
     P_e= 2T\sum_{\si=\pm1}\int\frac{d^3k}{(2\pi)^3}\ln\left[1+e^{-\left( k-\si\mu_e\right)/T}\right] \, .
 \end{equation}
The relative contribution of the thermal pressure \mbox{$P_\mathrm{th}=P(T)-P_\mathrm{cold}$}, where $P(T)$ is the sum of the finite temperature nucleon, electron, photon, and meson pressure, and $P_\mathrm{cold}$ is the sum of all pressure contributions at vanishing temperature, is plotted in Fig.~\ref{fig:PfinteT} for temperatures of $T=10$ and $T=50$ MeV.
 For comparison, see the results in Ref.~\cite{Raithel:2021hye}.
At low baryon densities, the thermal pressure at $T=10$ MeV is of the same order as the cold pressure, whereas for $T=50$ MeV the thermal pressure exceeds the cold pressure by an order of magnitude. For all models, the relative contribution of the thermal pressure becomes negligible at high baryon densities. QMC-RMF4 (purple line) shows the biggest relative thermal pressure at low densities, owing to its low $T=0$ pressure at low densities. 

We can now compute the expectation values of the meson fields by calculating the extrema of the pressure with respect to the meson condensates,
\begin{align}\label{eq:EOM}
\frac{\partial P}{\partial\langle\sigma\rangle}=0 \, , \qquad \frac{\partial P}{\partial\langle\omega_0\rangle}=0 \, , \qquad \frac{\partial P}{\partial\langle\rho_{03}\rangle}=0 \, .
\end{align}
This yields the following equations for the mean fields:
\begin{align}\label{eq:mesEOM}
     m_{\sigma}^2 \sev &=g_{\sigma} \left( n_{s n} + n_{s p} \right)- b M g_{\sigma}^3 \sev ^2 -c g_{\sigma}^4 \sev^3  \, , \nonumber \\
    m_{\omega}^2 \wev &=g_{\omega} \left(n_n+n_p\right) - 2b_1 g_{\rho}^2 \rev^2 \wev\, ,\nonumber \\
    m_{\rho}^2 \rev &=\frac{1}{2} g_{\rho} \left(n_p-n_n\right) - 2b_1 g_{\rho}^2 \rev \wev^2\, .
\end{align} 

The scalar densities for neutrons/protons are given by
\begin{equation}
  n_{si}= 2  \int\frac{d^3k}{(2\pi)^3}\frac{M_*}{E_i}\frac{1}{1+e^{(E_i-\mu_i)/T}} \, ,
\end{equation}
which reduces to
\begin{equation}
  n_{si}(T=0)=\frac{M_*}{2\pi^2}\left[k_{Fi}E_{Fi}^*-M_*^2\ln\frac{k_{Fi}+E_{Fi}^*}{M_*} \right] \, ,
\end{equation} in the $T=0$ limit. 
$M_*$ and $E_{Fi}^*$ are defined in Eq.~(\ref{eq:disprel}) and Eq.~(\ref{eq:estar}). The neutron/proton baryon  densities $n_{Bi}$ are given by
\begin{equation}
   n_{Bi}=2\int\frac{d^3k}{(2\pi)^3}\frac{1}{e^{(E_i-\mu_i)/T}+1} \, ,
\end{equation}
which reduces to the expression given in Eq.~(\ref{eq:nb}) in the zero temperature limit. The mean-field equations can now be solved numerically for a given temperature and chemical potentials for all particles.

\section{Parameterizing the $\chiPT$ Uncertainty Band}
\label{appx: param_chipt_band}
In Table \ref{tab:gcr_params} we give the values of $a$, $\alpha$, $b$, and $\beta$ that are used in Eq.~\eqn{eq:GCR} to generate the representatives of \chiEFT\ neutron matter to which our four RMFTs were fitted. 
In our approach these have no intrinsic significance: they are just an intermediate step in obtaining RMFTs that are consistent with \chiEFT\ results for neutron matter.

\begin{table}[h]
\begin{tabular}{l@{\quad}cccc}
\hline
Name & $a$ & $\alpha$ & $b$ & $\beta$ \\[-0.3ex]
   Unit & [MeV] &  & [MeV] &  \\ 
\hline
QMC-RMF1 & 14.996 & 0.628 & 1.854 & 3.039 \\
QMC-RMF2 & 14.638 & 0.526 & 1.948 & 3.186 \\
QMC-RMF3 & 17.796 & 0.847 & 0.296 & 5.218 \\
QMC-RMF4 & 13.758 & 0.389 & 0.824 & 4.844 \\
\hline
\end{tabular}
\caption{
The values of $a$, $\alpha$, $b$, and $\beta$ that (see Sec.~\ref{sec:fitting}) were used to obtain RMFTs that are
compatible with \chiEFT.
}
\label{tab:gcr_params}
\end{table}

\section{Asymptotic Speed of Sound}
\label{appx:csasym}
\label{appx:rmf_stability}
In this appendix we analytically compute the speed of sound for our class of RMFTs at asymptotic densities. In order to simplify the calculation, we restrict ourselves to isospin-symmetric matter, where the $\rho$ meson does not acquire a mean-field value. For large densities, the equations of motions in Eqs.~(\ref{eq:mesEOM}) simplify to 
\begin{align}
    m_\s^2\sev+bM g_{\s}^3\sev^2+cg_\s^4\sev^3&=g_\s \frac{\meff}{\pi^2}k_F^2 \, , \\
    \wev&=\frac{g_\om}{m_\om^2}\frac{2k_F^3}{3\pi^2} \, ,
\end{align}
where we used the fact that the effective mass decreases with density, which yields $n_s\approx\meff k_F^2 /\pi^2$ and $n_B$ is given by $n_B=2k_F^3/3\pi^2$. Note that neutrons and protons have identical magnitudes of the Fermi momentum $k_F$, leading to a factor of $2$ in $n_B$ and $n_s$. Solving the cubic equation for $\sev$ and only keeping the highest order term in $k_F$ yields
\begin{equation}
    \sev=\left(\frac{\meff}{cg_\s^3}k_F^2\right)^\frac{1}{3} \, ,
\end{equation}
therefore the mesonic pressure is dominated by the $\wev$-contribution and is given by
\begin{equation}
    P_M\approx\frac{2g_\om^2}{9m_\om^2}\frac{k_F^6}{\pi^4} \, .
\end{equation}
The nucleonic pressure, given in Eq.~(\ref{eq:Pnuc}), reduces to
\begin{equation}
    P_N\approx\frac{k_F^4}{6\pi^2}\, ,
\end{equation}
for $k_F\to\infty$ and is a subleading contribution to the total pressure. Given that the electron density is equal to the proton density and the Fermi momenta are therefore identical, we can conclude that the asymptotic pressure is given by the mesonic contribution $P_M$.
The energy density can be computed from the thermodynamic relationship $\varepsilon=-P+\mu_B n_B$. At $T=0$, $\mu_B-g_\om\wev=\sqrt{k_F^2+\meff^2}\approx k_F$, therefore at leading order we find that $\mu_B\approx\frac{g_\om^2}{m_\om^2}\frac{2k_F^3}{3\pi^2}$, which yields
\begin{equation}
    \varepsilon\approx\frac{2g_\om^2}{9m_\om^2}\frac{k_F^6}{\pi^4} = P_M\, .
\end{equation}
Finally, we compute the speed of sound at asymptotically high densities via
\begin{equation}
  c_s^2=\frac{\partial P}{\partial\varepsilon} = 1  \, ,
\end{equation}
which is independent of any couplings and numerically confirmed in Fig.~\ref{fig:speed_of_sound}.
\clearpage

\bibliography{chiPT_RMF}

\end{document}